\documentclass[fleqn,usenatbib]{mnras}

\usepackage{newtxtext,newtxmath}
\usepackage[T1]{fontenc}
\usepackage{ae,aecompl}

\usepackage{natbib}

\usepackage{graphicx}	
\usepackage{amsmath}
\usepackage{hyperref}
\usepackage{listings}
\usepackage{color}
\usepackage{float}
\usepackage{xspace} 
\usepackage{longtable}

\newcommand{\kepler}{\textsl{Kepler}\xspace}

\newcommand{\gaia}{\textsl{Gaia}\xspace}

\title[The solar light curve]{The Solar Benchmark: Rotational Modulation of the Sun Reconstructed from Archival Sunspot Records}

\author[Morris et al.]{Brett M. Morris,$^{1}$\thanks{E-mail: morrisbrettm@gmail.com }
James R.A. Davenport,$^{1, 2}$
Helen A.C. Giles,$^{3}$
Leslie Hebb,$^{4}$ \newauthor
Suzanne L. Hawley,$^{1}$
Ruth Angus,$^{5}$
Peter A. Gilman,$^{6}$
Eric Agol$^{1}$\\
$^{1}$Astronomy Department, University of Washington, Seattle, WA 98195, USA\\
$^{2}$DIRAC Fellow\\
$^{3}$Observatoire de Gen\`{e}ve, Universit\'{e} de Gen\`{e}ve, Chemin des Maillettes 51, 1290 Versoix, Switzerland \\
$^{4}$Physics Department, Hobart and William Smith Colleges, Geneva, NY 14456, USA\\
$^{5}$ American Museum of Natural History, New York, NY, USA\\
$^{6}$High Altitude Observatory, National Center for Atmospheric Research, Boulder, Colorado, USA\\
}

\date{Accepted XXX. Received YYY; in original form ZZZ}

\pubyear{2017}

\begin{document}
\label{firstpage}
\pagerange{\pageref{firstpage}--\pageref{lastpage}}
\maketitle

\begin{abstract}
We use archival daily spot coverage measurements from Howard et al.\ (1984) to study the rotational modulation of the Sun as though it were a distant star. A quasi-periodic Gaussian process measures the solar rotation period $P_\mathrm{rot} = 26.3 \pm 0.1$ days, and activity cycle period $P_\mathrm{cyc} = 10.7 \pm 0.3$ years. We attempt to search for evidence of differential rotation in variations of the apparent rotation period throughout the activity cycle and do not detect a clear signal of differential rotation, consistent with the null results of the hare-and-hounds exercise of Aigrain et al.\ (2015). The full reconstructed solar light curve is available online.
\end{abstract}


\section{Introduction}

For decades astronomers have endeavoured to study the ``Sun as a star'', measuring properties of the Sun that we typically measure on distant stars, with the goal of putting the Sun into context \citep[e.g.:][]{Livingston1991, Tayler1996, Chaplin2004, Livingston2007, Hall2009, Bertello2012, Hall2015, Egeland2017}. These efforts are valuable, for example, for understanding the Sun's activity through time, by observing Sun-like stars of different ages or at different phases in their activity cycles.

We are entering a new era for the study of rotational modulation of stars. \kepler has measured rotational modulation of tens of thousands of stars for four consecutive years, and K2 has measured rotation periods for many more stars, albeit over a shorter baseline. TESS will measure precision light curves for bright nearby stars, for a maximum duration of 355 consecutive days near the ecliptic poles in the primary mission \citep{Ricker2014, Sullivan2015}. \gaia will measure rotation periods for $>10^5$ stars \citep[see, e.g.][]{Lanzafame2018}. ESA's PLAnetarty Transits and Oscillations (PLATO) mission may observe targets for up to 8 years \citep{Rauer2014}, potentially allowing us to probe variations in the stellar rotational modulation of stars as a function of phase in their activity cycles. Having a solar benchmark light curve to compare these future, long-term light curves will be an important data product for the community.  

\citet{Morris2018b} developed tools for measuring the apparent stellar centroid offsets due to starspots that affect \gaia astrometry. In particular, a framework was developed for reconstructing archival spot maps of the Sun using the Mount Wilson Observatory (MWO) spot coverage catalog published in \citet{Howard1984}. The MWO spot catalog is a digitized representation of ``white light'' photographic plate images of the solar disk taken from 1917-1985, denoting the apparent positions (latitude and longitude) and areas of penumbrae in sunspot groups. In this work, we use the same software and spot coverage archive as \citet{Morris2018b}, to reconstruct artificial time-series photometry of the Sun with one-day cadence. 

In Section~\ref{sec:lc} we introduce our approximation of the solar rotational light curve, and measure its properties as though it were a distant star. We will then recover several properties of the Sun using the reconstructed light curve. First and foremost we seek to recover the solar rotation period and activity cycle period, which are 25-34 days and 10.9 years respectively \citep{Howe2000, Hathaway2015}. 

We also follow the technique of \citet{Giles2017} to estimate the sunspot lifetimes. High resolution observations of sunspots show that they have lifetimes ranging from from hours to months \citep{Solanki2003}. There is a roughly linear relationship between active region areas and their lifetimes, as described by \citet{Gnevyshev1938} and \citet{Waldmeier1955} \cite[see also, for example, ][]{Petrovay1997}. 

The broad range of possible rotation periods for the Sun is the result of differential rotation -- the Sun rotates faster at the equator than at the poles \citep{Miesch2005}. The pursuit to detect differential rotation from photometric rotational modulation of Sun-like stars in \kepler light curves has proven very difficult \citep{Aigrain2015}. Setting the perils aside, we will naively attempt to search for differential rotation by its effect on the solar light curve in Section~\ref{sec:dr}.

\section{The Solar Light Curve} \label{sec:lc}

\subsection{Constructing the light curve}

\begin{figure*}
    \centering
    \includegraphics[scale=0.9]{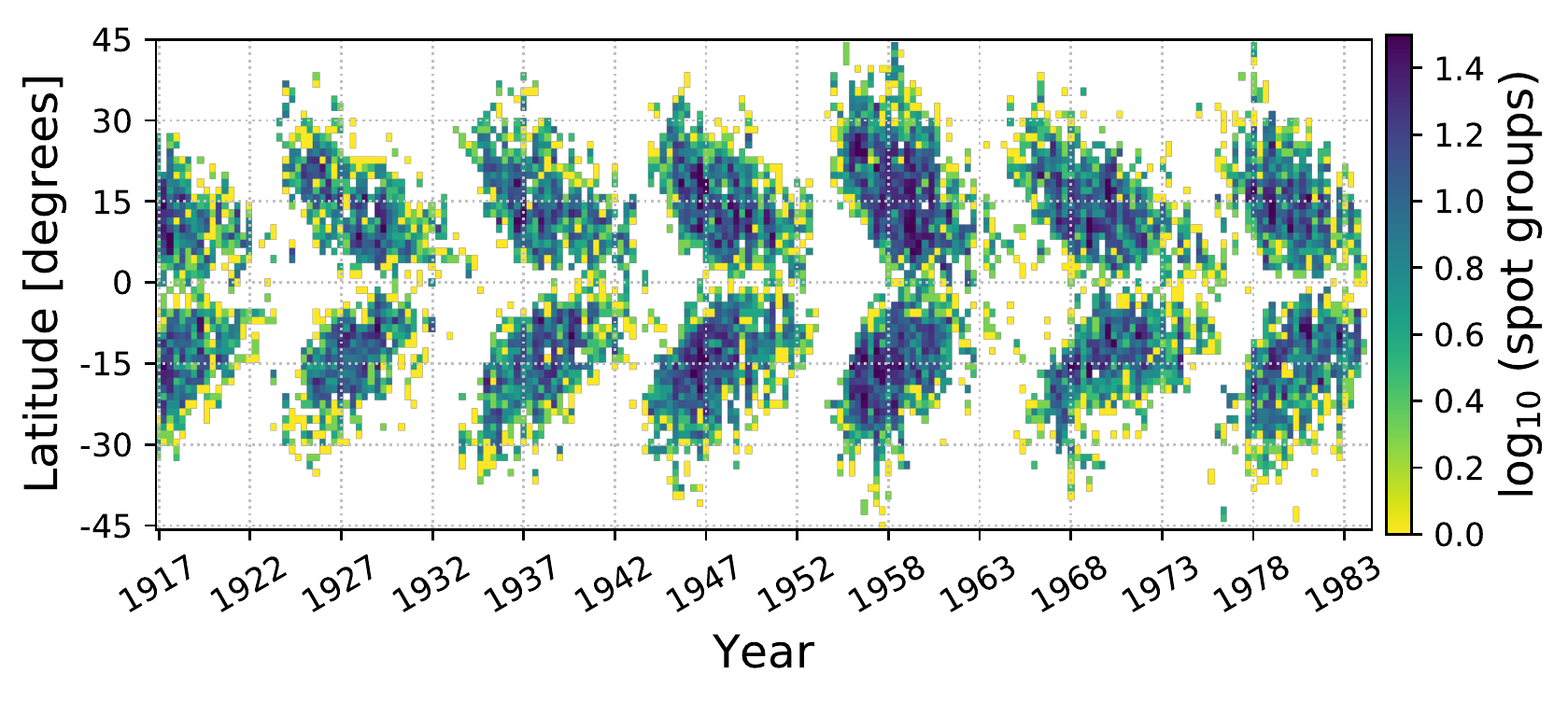}
    \caption{Butterfly diagram (after \citealt{Maunder1904}) showing spot density as a function of time and solar latitude with the spot archive of \citet{Howard1984}.}
    \label{fig:butterfly}
\end{figure*}

\begin{figure*}
    \centering
    \includegraphics[scale=0.9]{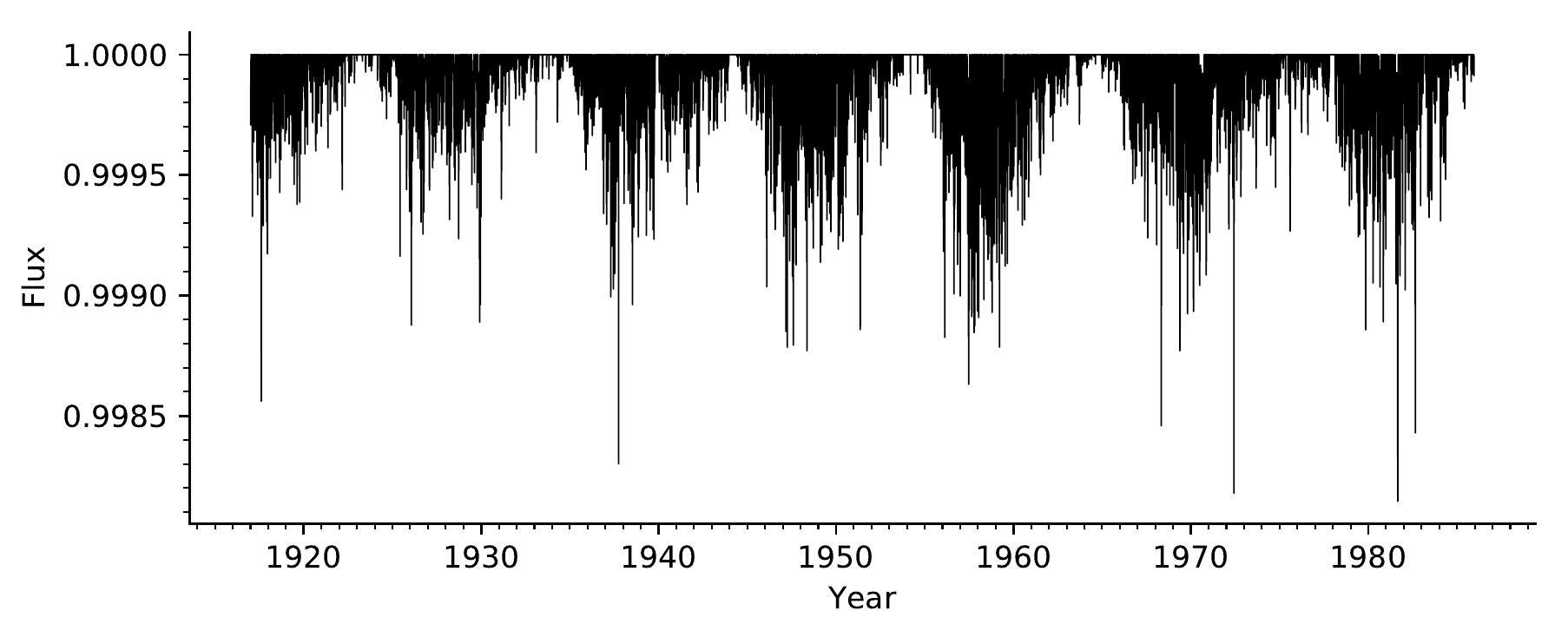}
    \caption{Reconstructed solar light curve from the spot area coverage archive of \citet{Howard1984}. The standard deviation of the full light curve is 150 ppm. The mean flux is 80 ppm less than the maximum flux. See Figure~\ref{fig:cycle19} for a close-up view of one cycle and further descript1ion.}
    \label{fig:lc}
\end{figure*}

\begin{figure*}
    \centering
    \includegraphics[scale=0.9]{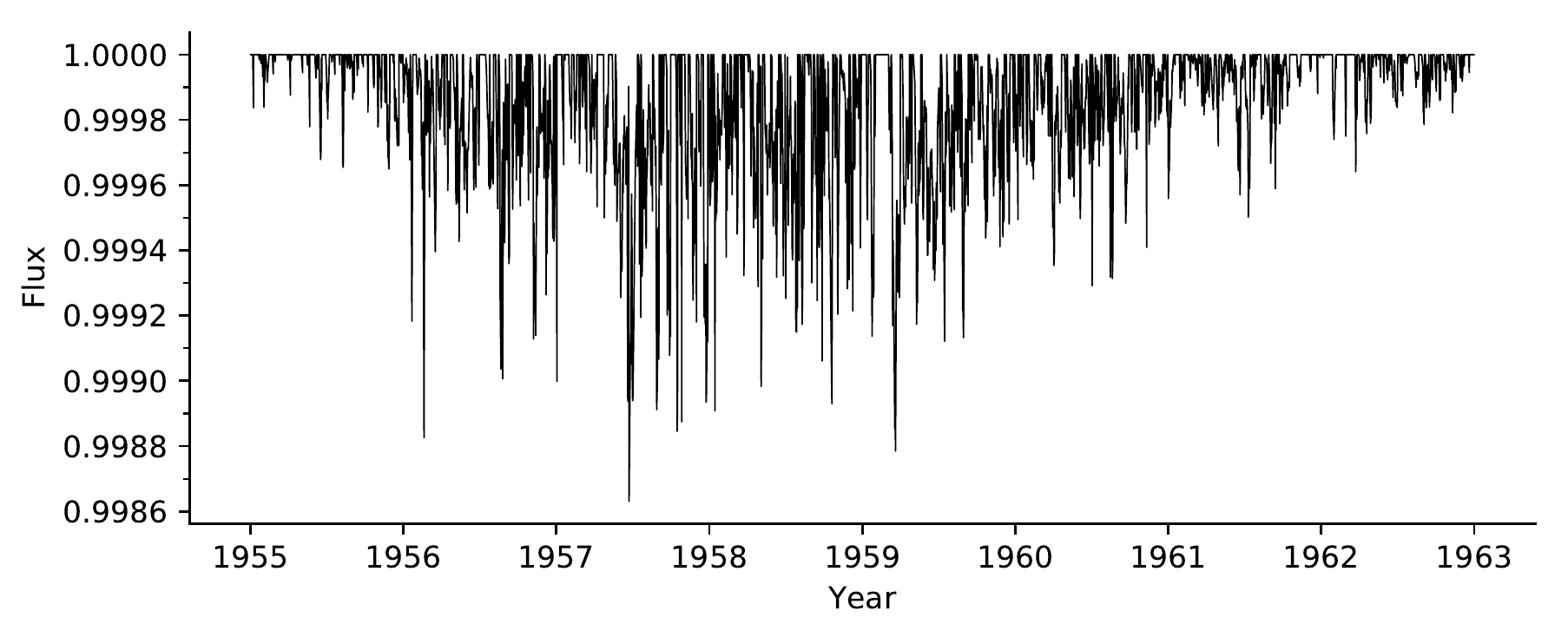}
    \caption{Reconstructed solar light curve zoomed into cycle 19 to show fine structure. This light curve is unlike \kepler light curves for several reasons: our reconstruction has no photon noise, no p-mode oscillations, no granulation ``flicker'', and no instrumental artifacts. In addition, unlike \kepler targets, we know the true unspotted flux of the Sun in these reconstructed light curves (i.e. we know when the Sun was truly spotless), so the light curve has a maximum of unity, rather than a median of unity.}
    \label{fig:cycle19}
\end{figure*}

As in \citet{Morris2018b}, we integrate the total flux of the unspotted, limb-darkened Sun,
\begin{equation}
F_{\odot, \mathrm{unspotted}} = \int_{0}^{R_\star} 2 \pi r \, I(r) dr,
\end{equation}
where $I(r)$ is a quadratic limb-darkening law and $r$ is in units of angle, so that $2\pi rdr$ is solid angle.

We define Cartesian sky-plane coordinates $(x,y)$, with the origin placed at the center of the star, $\hat{x}$  aligned with the stellar equator, and $\hat{y}$ aligned with the stellar rotation axis. We describe each starspot with an ellipse with centroid ${\bf r}_i = (x_i, y_i)$, and $r_i = \vert {\bf r}_i \vert$.  We can compute the negative flux contribution from each spot by computing the approximate spot area and contrast. A circular spot will be foreshortened near the stellar limb. The foreshortened circular spot can be approximated with an ellipse with semi-major axis $R_{\mathrm{spot}}$ and semi-minor axis $R_{\mathrm{spot}} \sqrt{1 - (r_i/R_\odot)^2}$. 

Since these spots are small compared to the solar radius ($R_{\mathrm{spot}}/R_\odot < 0.1$), we adopt one limb-darkened contrast for the entire spot, $c_{ld} = (1-c) I(r)$, where $c$ is the flux contrast in the spot relative to the photosphere flux. 

The integrated spot flux is 
\begin{equation}
F_{\mathrm{spot}, i} = - \pi R_{\mathrm{spot}}^2 c_{ld} \sqrt{1 - (r_i/R_\odot)^2}, 
\end{equation}
and accounting for all $N$ spots, the spotted flux of the star is
\begin{equation}
F_{\odot, \mathrm{spotted}} = F_{\odot, \mathrm{unspotted}}  + \sum_{i=1}^{N} F_{\mathrm{spot}, i}.
\end{equation}
This approximation is valid for spots that are small compared to the solar radius, or small compared to the scale of limb-darkening variation across the solar disk.

The spot group coverage catalog of \citet{Howard1984} describes the daily areas and positions of sunspot groups from 1917-1985, see Figure~\ref{fig:butterfly}. We approximate each spot group as a single circular spot with the area of the entire spot group. We fix the spot contrast in the \kepler band at $c = 1 - I_\mathrm{spot}/I_\mathrm{phot} = 0.7$, which is the mean sunspot intensity averaged over the penumbra and umbra, assuming their typical penumbra covers a factor of 5 more area than the umbra \citep{Solanki2003}. 

The reconstructed solar light curve is shown in Figures~\ref{fig:lc} and \ref{fig:cycle19}. This very long-term view of the solar light curve shows periods of high variance separated by relatively quiet times, corresponding to the phase in the activity cycle. During solar maximum, there can be as many as 14 spot groups on the visible hemisphere of the Sun at once, leading to typical dips in flux of order $\sim 500$ ppm. Near solar minimum, the spotless solar surface had no spot groups, and we have filled in those dates with no spot group entries with flux equal to unity. 

The full reconstructed solar light curve is available online \citep{solarlightcurve}\footnote{\url{https://doi.org/10.5281/zenodo.1476637}}. 

\subsection{Constraining the Effects of Faculae}

The Mount Wilson Observatory sunspot catalog only tracked the positions and areas of dark sunspots, but did not measure the positions or sizes of faculae, which are small \textit{bright} regions of concentrated magnetic flux. We reconstruct the solar light curve due to facular brightening using the same technique as in the previous section, but this time using faculae positions and areas from the Greenwich Photo-Heliographic Plate archive, digitized in 1999 by the NOAA Environmental Data Rescue Program, which provides facular positions and areas.

Unlike the starspots, we do not choose a fixed contrast for the faculae, since facular intensity varies as a function of position on the Sun. Therefore we compute a contrast for each facula individually given their position according to
\begin{equation}
\Delta T_{\rm fac} = 250.9 - 407.7 \mu + 190.9 \mu^2,
\end{equation}
where $\Delta T_{\rm fac}$ is the temperature excess of the faculae relative to the local photosphere, $\mu = \cos \theta$, and $\theta$ is the angle between the stellar surface normal and the observer's line of sight \citep{Meunier2010,Dumusque2014}. The contrast of each facula is thus the integrated blackbody flux with the photospheric temperature plus the temperature excess, normalized by the blackbody flux with the temperature of the photosphere (5777 K). We integrate the blackbodies over the \kepler bandpass, but the choice of bandpass has little effect on the results (see, for example, Figure 2 of \citealt{Morris2018b}).

\begin{figure}
  \centering
  \includegraphics[scale=0.8]{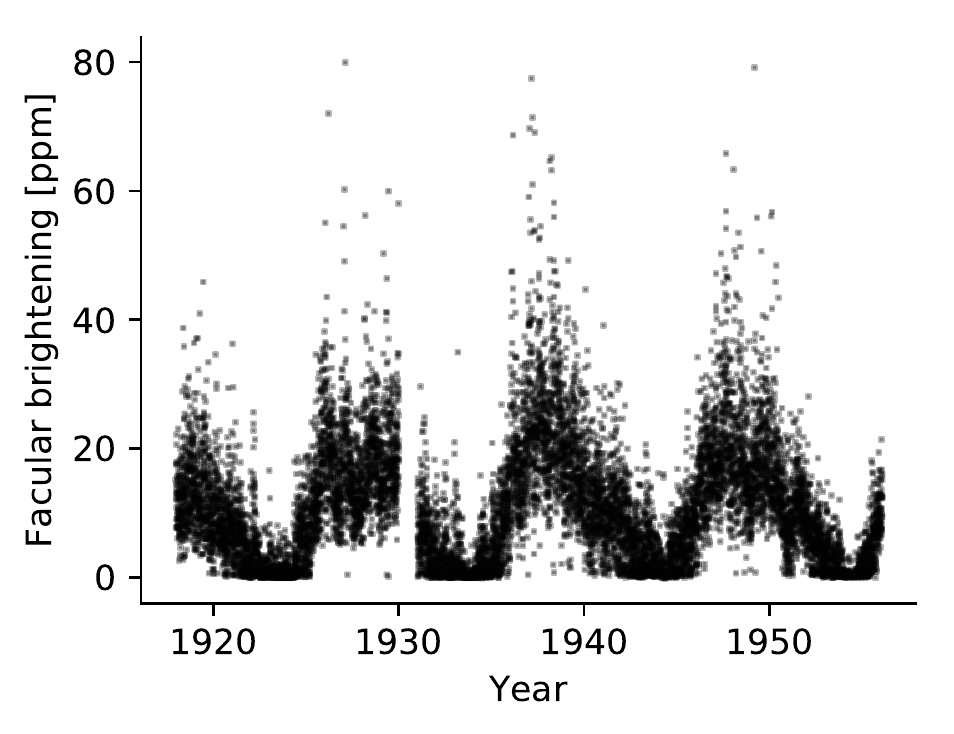}
  \caption{Facular brightening of the Sun from the Greenwich Photo-Heliographic archive spanning three activity cycles. Note that the scale of facular brightening in the \kepler band is small ($\lesssim 20$ ppm) compared to the darkening due to sunspots ($\lesssim 200$ ppm).}
  \label{fig:faculae}
\end{figure}

The resulting light curve of excess solar flux due to faculae is shown in Figure~\ref{fig:faculae}.  Typical brightening in the \kepler band due to faculae is small ($\lesssim 20$ ppm) compared to the darkening due to sunspots ($\lesssim 200$ ppm). Despite their large relative area coverage compared to starspots, the typical facular intensity contrast ($c \sim 1.05$) is relatively small compared with sunspots ($c \sim 0.7$), so we expect spots to dominate the rotational modulation of the Sun in the \kepler band, in agreement with \citet{Shapiro2016}, for example. 

The dominance of sunspots over faculae in the rotational light curve of the Sun is not to be confused with the fact that the Sun is considered ``faculae dominated''  on timescales of the activity cycle. That is, near solar maximum the Sun is bolometrically brighter than it is at solar minimum \citep{Solanki2013}. What we refer to as the solar light curve in this work is \textit{not} the bolometric flux of the Sun, rather it is the flux integrated over a bandpass like those of \kepler, TESS, or \gaia. As such, we choose to ignore the effects of faculae in the remainder of this work, since spots dominate the rotational modulation, which is our primary focus. 

\subsection{Measuring the solar rotation and activity cycle periods} \label{sec:gp}

An astronomer's first instinct is likely to measure periodicities upon seeing a light curve such as Figure~\ref{sec:lc}. In this section, we examine the autocorrelation function and Lomb-Scargle periodogram of the solar light curve to establish benchmark measurements of the rotation and activity cycle periods.

\subsubsection{Gaussian Process regression}

\begin{figure*}
    \centering
    \includegraphics[scale=0.85]{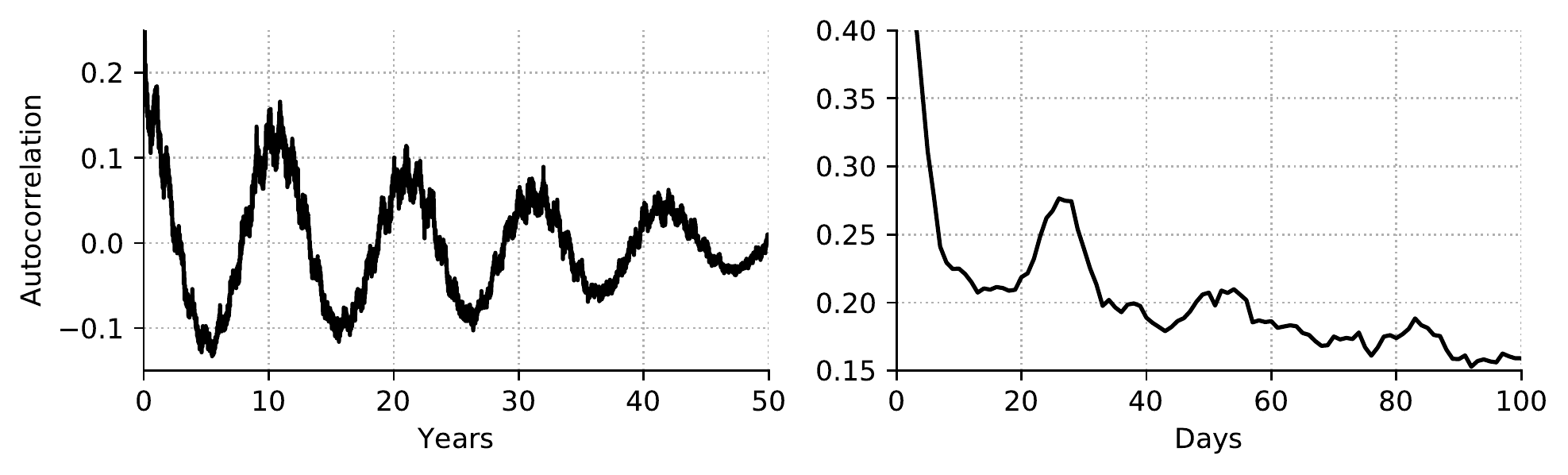}
    \includegraphics[scale=0.85]{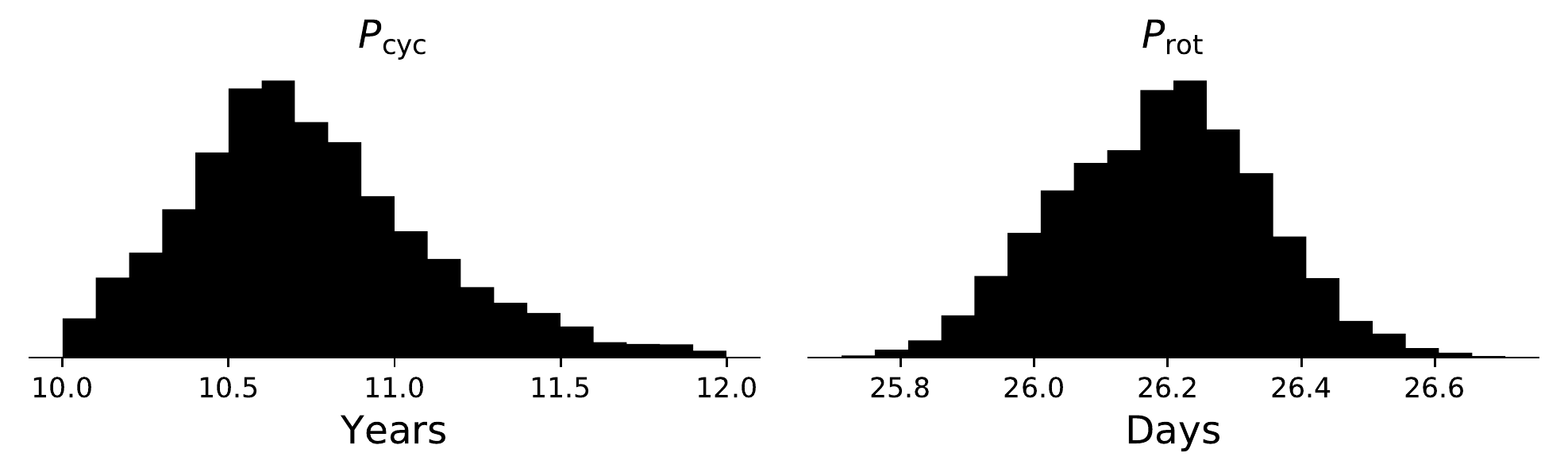}
    \caption{{\it Upper left}: long-term autocorrelation function of the solar light curve in Figure~\ref{fig:lc}. The first peak is $\sim10.5$ years. {\it Upper right}: short-term signals in the autocorrelation function of the solar light curve, with a peak at 26 d. This estimate of the rotation period is approximately consistent with the rotation period at the active latitudes \citep{Howe2000}. See Section~\ref{sec:gp} for a more robust rotation period measurement using Gaussian process regression with a quasi-periodic kernel. {\it Lower}: Posterior distributions for the the magnetic activity cycle period $P_\mathrm{cyc}$ and the solar rotation period $P_\mathrm{rot}$ measured with a quasi-periodic Gaussian process regression to the synthetic photometry from 1917-1985 (see Figure~\ref{fig:lc}). We measure $P_\mathrm{rot}=26.32 \pm 0.14$ days and $P_\mathrm{cyc} = 10.61 \pm 0.23$ years.}
    \label{fig:acf}
\end{figure*}

The autocorrelation function of the solar light curve is shown in Figure~\ref{fig:acf}. There is short-term variation peaking at 26 days -- approximately the rotation period of the Sun at the photosphere near the active latitudes \citep{Howe2000}. There is also a long-term decaying cosine-shaped correlation with its first peak at $10.6$ years, corresponding to the magnetic activity cycle period of $\sim 11$ years \citep{Hathaway2015}. Finally, there is a cosine-shaped correlation with a period of 365 d, corresponding to the orbital period of the Earth. This systematic crops up because the Earth's orbit is inclined with respect to the solar equator by 7.25$^\circ$ \citep{meeus}, causing starspots to drift slightly towards and away from the solar equator throughout each year,  injecting a small correlated signal into the reconstructed light curve. 

For a more rigorous measurement of the solar rotation and activity cycle period, we model the light curve with a quasi-periodic Gaussian process with a kernel of the form: 
\begin{equation}
\begin{split}
    k(\tau) = a_0 e^{-c_0 \tau} \cos\left(\frac{2\pi \tau}{P_\mathrm{cyc}}\right) +\\
    a_1 \cos\left(\frac{2\pi \tau}{P_\oplus}\right) + \\ 
    a_2 e^{-c_2 \tau} \left[\cos\left(\frac{2\pi \tau}{P_\mathrm{rot}}\right) + 1 \right] \\ 
\end{split}
\end{equation}
$\tau$ is the difference in times (units of days). The exponential term allows for deviations from a perfectly periodic activity cycle signal with decay timescale $c_0 > 0$. $P_\mathrm{rot}$ is the rotation period and $P_\mathrm{cyc}$ is the activity cycle period. $P_\oplus$ is the orbital period of the Earth, which imprints itself on these data because the Earth's inclination with respect to the solar equator gives rise to a periodic systematic shift in the positions of sunspots. We fit for $a_0, a_1, a_2, c_0, c_2, P_\mathrm{rot}, P_\mathrm{cyc}$ using Markov Chain Monte Carlo via \texttt{emcee} with \texttt{celerite} \citep{Foreman-Mackey2013, Foreman-Mackey2017}. We measure $P_\mathrm{rot}=26.3 \pm 0.1$ d  -- see the posterior distributions in Figure~\ref{fig:acf}. We note that this is consistent with the asteroseismic rotation period of the solar photosphere at $\sim 15^\circ$ latitude \citep{Howe2000} -- as one would hope, it seems the quasi-periodic Gaussian process properly recovers the rotation period at the active latitudes where the most spots are emerging. Thus, at high enough S/N, a light curve will show the rotation period at the active latitudes, rather than the equatorial rotation period, as has been potentially observed in tidally synchronized binaries \citep[see e.g.][]{Lurie2017}. 

We also measure activity cycle period $P_\mathrm{cyc} = 10.61 \pm 0.23$ years. This is consistent with canonical cycle period measured by taking the dates of the minima of cycle 1 and cycle 23 and dividing by 22, yielding an average cycle period of 10.9 years (131.7 months, \citealt{Hathaway2015}). 

\begin{figure*}
    \centering
    \includegraphics[scale=0.8]{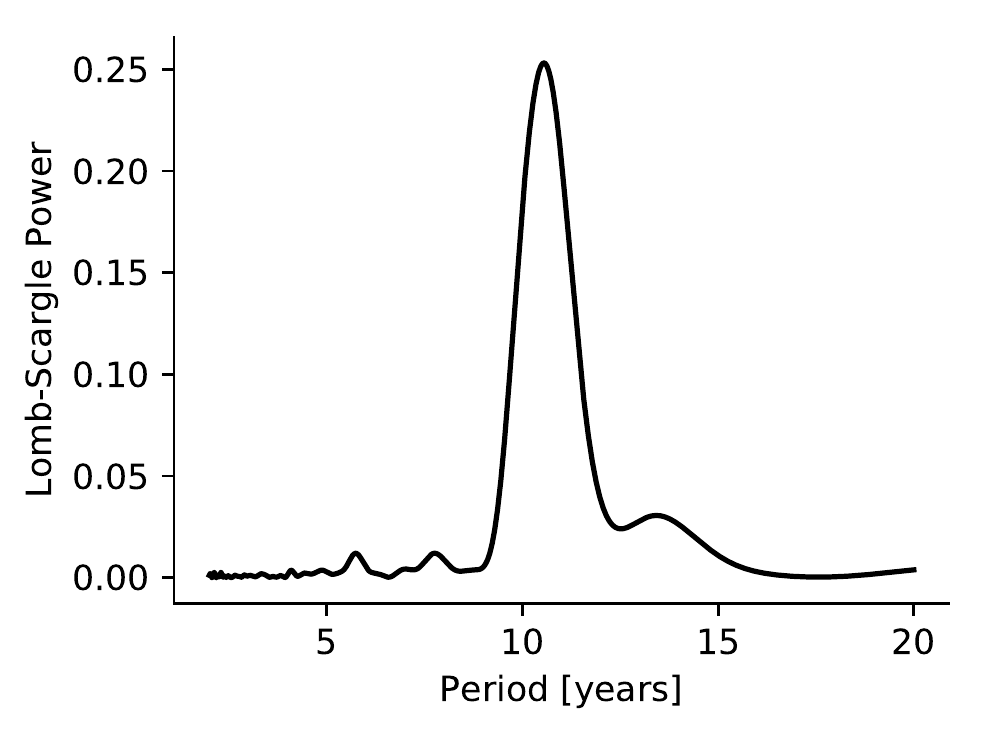}
    \includegraphics[scale=0.8]{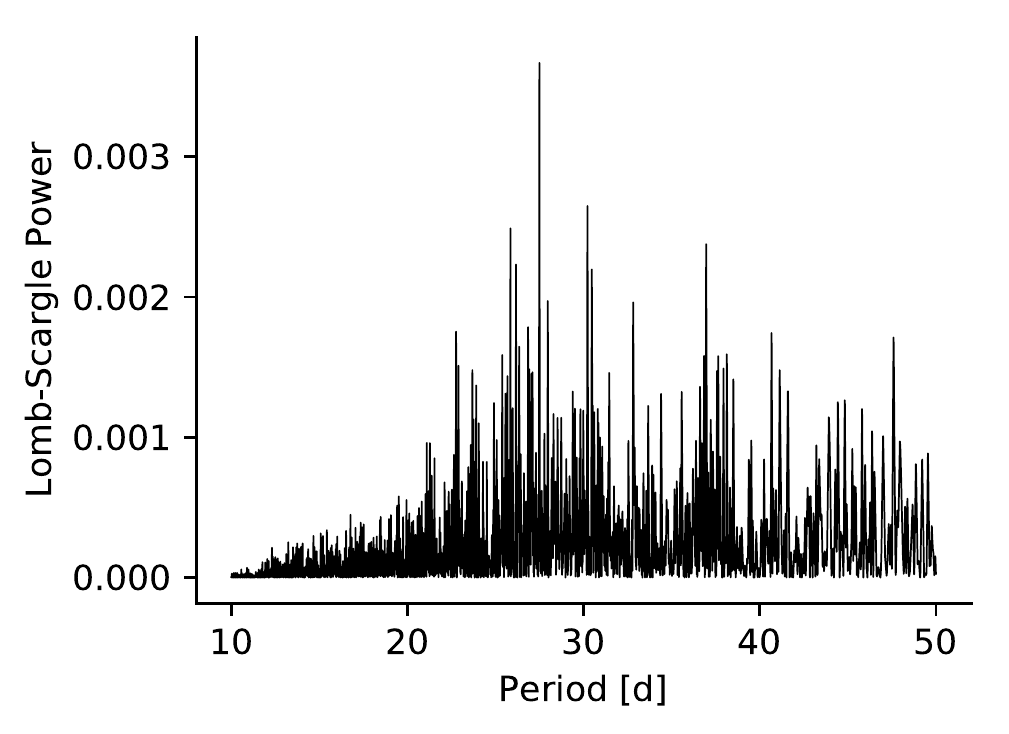}
    \caption{Lomb-Scargle periodogram of the solar light curve in Figure~\ref{fig:lc} on long ({\it left}) and short ({\it right}) timescales. Like the autocorrelation function in Figure~\ref{fig:acf}, the Lomb-Scargle periodogram has a peak at the oft-quoted activity cycle period of 11 years, and the rotational peak at 27 days.}
    \label{fig:ls}
\end{figure*}

\subsubsection{Lomb-Scargle Periodogram}

Next, we use the Lomb-Scargle (LS) periodogram to compare its ability to pick out the quasi-periodic peak -- see Figure~\ref{fig:ls}. The dominant period is 27 d, just longer than the rotation period measured by Gaussian process regression in the previous section. The difference in rotation periods measured with each technique is an artifact of the intrinsically quasi-periodic nature of the Gaussian process kernel in the previous section, and the strict periodicity enforced by the LS periodogram. In addition, the uncertainty in the periodicity measured with the LS periodogram is not well defined, so it is not possible to do a robust comparison between the LS and Gaussian process period measurements. Turning to longer periods, the activity cycle peak is prominent at 10.6 years -- this result is consistent with the Gaussian process regression measurement.

We prefer the value from the quasi-periodic Gaussian process analysis for the apparent rotation period rather than the LS period because: (1) we know from high resolution observations that there's more than one frequency at play -- for example, starspots emerge and decay on timescales similar to the stellar rotation period; and (2) the Gaussian process regression provides us with robust uncertainties on the apparent rotation period. For these reasons, we encourage observers of distant stars to consider using Gaussian process regression over the LS periodogram when searching for the rotation period at the mean active latitudes \citep{VanderPlas2018}.

\subsection{Measuring active region evolution timescales}

The autocorrelation function in Figure~\ref{fig:acf} has a peak at the rotation period of the star, and smaller peaks at integer multiples of the rotation period with decreasing amplitudes. \citet{Giles2017} developed a technique for measuring active region evolution timescales by modeling the autocorrelation functions of active stars with a underdamped simple harmonic oscillator (uSHO), which we apply here to the autocorrelation function of the reconstructed solar light curve to estimate active region lifetimes. 

We attempt to measure the sunspot lifetimes from the autocorrelation function. However the form of the autocorrelation function (Figure~\ref{fig:acf}) is slightly different to those seen in \citet{Giles2017}, which typically follow the pattern of an underdamped simple harmonic oscillator. In Figure~\ref{fig:acf} there is an additional decreasing trend which causes the subsequent peaks to be significantly lower than the central peak, which arises from the much longer timescale activity cycle pattern. This effect is persistent whether we generate the autocorrelation function for the light curve as a whole, or cut it up into smaller portions and combine the autocorrelation functions. 

Although the uSHO fits were unsuccessful, we can still make some qualitative statements from inspection of the autocorrelation function at short lags. The signal of rotation peaking at 26 days has repeated aliases at twice and possibly at three times the rotation period, each with diminished amplitude, before the aliases of the rotation signal appear to decay away at large lags ($\gtrsim 3 P_\mathrm{rot}$). This suggests that the typical spot decay timescale is similar to the rotation period, and only occasional spots survive more than one or two solar rotations. This observation is in agreement with spatial resolved observations which show that the longest-lived sunspots live of order several rotations \citep{Pettit1951, Howe2000}, but most only survive for less than one rotation \citep{Petrovay1997}.

\section{Differential rotation} \label{sec:dr}

Many efforts have been made to quantify differential rotation in \kepler light curves of stars, most notably in \citet{Aigrain2015}, where several groups of observers attempted to measure the differential rotation rate in synthetic light curves. The authors found that there was little relation between the injected and recovered differential rotation rates, indicating that \kepler detections of solar-like differential rotation ought to be treated with caution. 

In this Section, we set out to mimic this perilous exercise by measuring the solar rotation period in consecutive one year bins, using the quasi-periodic Gaussian process technique that we used in Section~\ref{sec:gp} to measure the rotation period of the full light curve. We choose one year bins so that there is sufficiently long baseline to get a fit for the period, but the duration is short compared to the activity cycle period (11 years). If the rotational modulation contains the signature of differential rotation, we expect to find that the apparent rotation period changes slightly from one year to the next, as spots emerge at different latitudes throughout the activity cycle, and due to differential rotation, the spots rotate with slightly different periods. 

Ideally, we would observe that at the beginning of each activity cycle, the spots emerge at high latitudes and therefore the apparent rotation period is long. Then as the activity cycle progresses, spots emerge at lower latitudes, revealing shorter rotation periods. 

\begin{figure}
    \centering
    \includegraphics[scale=0.9]{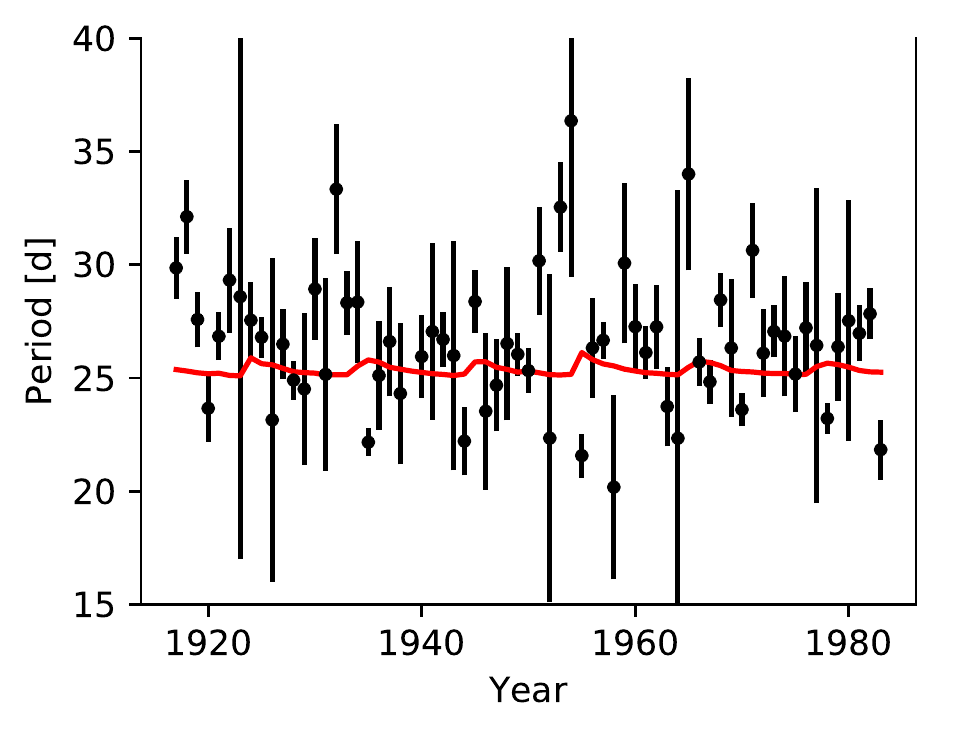}
    \caption{Measured solar rotational period inferred from quasi-periodic Gaussian process regression to one year-long bins of the solar light curve (black circles), compared with the rotation period at the mean area-weighted spot latitude averaged in one year bins (red curve). As spots emerge at different latitudes with differential rotation, we hoped to find that the rotation period varied from year to year with the phase of the activity cycle -- with spots emerging at high latitudes and long rotation periods, and the rotation period appearing to decrease as spots emerge closer to the solar equator. It appears that due to spots emerging at a broad range of latitudes at all phases of the activity cycle, the apparent rotation period remains largely constant, irrespective of the activity cycle phase.}
    \label{fig:period_cycle}
\end{figure}

The rotation period recovered from fitting the quasi-periodic Gaussian process to one-year bins of the solar light curve is shown in black points in Figure~\ref{fig:period_cycle}. The red curve shows the rotation period at the mean area-weighted spot latitude averaged into yearly bins, and shows the small differential rotation signal imparted by the activity cycle which we are attempting to measure. In practice, we observe a spread in measured rotation periods much larger than the variance due to the activity cycle, with similarly large uncertainties. Activity minima can be identified in this figure by the large uncertainties on the rotation period, when few spots are present to drive rotational modulation. In between these points of large uncertainties are intervals where the rotation period is measured more precisely, though it is roughly consistent with a 26.3 d rotation period throughout all phases of the activity cycle. Assuming the rotation period is 26.3 d throughout, the reduced $\tilde{\chi}^2 = 8$, indicating that the variance is indeed greater than expected for Gaussian-distributed errors. However, the stochastic nature of the measurements make it impossible to recover the true differential rotation rate from these rotation measurements. Therefore even at ``infinite'' signal-to-noise, we arrive at the same conclusion as \citet{Aigrain2015} -- measuring differential rotation shear from rotational modulation alone is a fraught exercise. 

\section{Discussion} \label{sec:discussion}

The detection of differential rotation from the solar light curve eludes us in Section~\ref{sec:dr}. One reason for this is made clear by the butterfly diagram in Figure~\ref{fig:butterfly} -- the distribution of spots within active latitudes of the Sun are broad; spots are distributed within $\pm 8^\circ$ of the mean ``active latitude'' at each phase of the activity cycle. The spots at multiple latitudes each contribute to the rotational modulation with their own rotation period, imprinting the mean rotation period on the light curve, rather than the specific rotation period at a high or low latitude. We are encouraged by recent work by \citet{Benomar2018} which may hold the key to measuring differential rotation from stellar photometry via asteroseismology for at least a small subset of stars. 

One limitation of this reconstruction approach is that the \citet{Howard1984} spot archive only cataloged spots within 60$^\circ$ longitude of the central solar meridian, meaning that spots on the limb were not logged. If spots near the limb were included in this time series, the overall flux trends might be smoother, and there would be fewer days with flux equal to unity. However, the net effect on the rotation period and activity cycle measurements is likely small, since spots on the limb are geometrically foreshortened, and due to the Wilson depression, they have smaller contrasts than spots at disk center \citep{Solanki1993}. 

\section{Conclusion}

We reconstructed a one day-cadence light curve of the Sun using the sunspot archive of \citet{Howard1984}. We compared the amplitude of variability due to dark sunspots to the amplitude of brightening from faculae, and found that the dark sunspots dominate the rotational modulation.

With the noise-free light curve, we measured rotation period and activity cycle period of the Sun with both the Lomb-Scargle periodogram and a quasi-periodic Gaussian process regression. The rotation periods and activity cycle periods measured with both techniques are consistent with the rotation period at the active latitudes, and the duration of a typical activity cycle. We showed that differential rotation cannot be detected even from this idealized reconstructed light curve.

\section*{Acknowledgements}

We gratefully acknowledge the following softoware packages which made this work possible: \texttt{astropy} \citep{Astropy2013, Astropy2018}, \texttt{ipython} \citep{ipython}, \texttt{numpy} \citep{VanDerWalt2011}, \texttt{scipy} \citep{scipy},  \texttt{matplotlib} \citep{matplotlib}, \texttt{celerite} \citep{Foreman-Mackey2017}, \texttt{gatspy} \citep{gatspy}, \texttt{emcee} \citep{Foreman-Mackey2013}. Some of this work has been carried out in the frame of the National Centre for Competence in Research `PlanetS' supported by the Swiss National Science Foundation (SNSF).

JRAD acknowledges support from the DIRAC Institute in the Department of Astronomy at the University of Washington. The DIRAC Institute is supported through generous gifts from the Charles and Lisa Simonyi Fund for Arts and Sciences, and the Washington Research Foundation. This research has made use of NASA's Astrophysics Data System.

\bibliographystyle{mnras}
\bibliography{bibliography}

\bsp	
\label{lastpage}
\end{document}